\documentclass[twocolumn,showpacs,aps, prl]{revtex4-2}%
\usepackage{amsfonts}
\usepackage{amsmath}
\usepackage{amssymb}
\usepackage{dcolumn}% Align table columns on decimal point
\usepackage{epsf, epsfig, graphics}
\usepackage{subfigure}
\usepackage{graphicx}
\usepackage{color}
\usepackage{comment}
\usepackage{epstopdf}

\begin{document}
\title{How do incorrect ligands help detect a correct ligand?}
\author{Yan-Ru Chen$^{1}$, Kwan-tai Leung$^{2}$, and Hsuan-Yi Chen$^{1,2}$ %\footnote{hschen@phy.ncu.edu.tw}
}
 \affiliation{$^{1}$Department of Physics and Center for Complex Systems, National Central University, Jhongli, 32001, Taiwan\\
 $^{2}$Institute of Physics, Academia Sinica, Taipei, 11529, Taiwan}	 				
%%%%%%%%%%%%%%%%%
 \begin{abstract}		
    Intrigued by the response of T cell receptors to the presence of a few agonist ligands, we propose a minimal model that can achieve similar performance.  
The model consists of a small cluster of immobile receptors that bind reversibly to two types (correct/incorrect) of ligands in the environment, with slightly weaker binding strength for the incorrect one. 
It features binding-state coupling between nearest-neighbor receptors, and receptors in the bound/free states are activated/deactivated by specific enzymes, with rates that allow kinetic proofreading.    
It is found that, for a range of binding-state coupling strengths, incorrect ligands alone cannot activate the receptors, but the binding of merely one correct ligand to a receptor is sufficient to promote the activation of other receptors via induced binding to incorrect ligands.
Both response time and signal amplification increase as the receptor binding-state coupling strength increases until it reaches an optimal range to achieve the most rapid and sensitive response.    
These results suggest a possible mechanism for a speedy and specific response of receptors to very few correct ligands in biological and artificial systems at the subcellular scale.
\end{abstract}	
 \date{\today}
 \maketitle 
 
 \newpage    	 
%%%%%%%%%%%%%%%    	   
 {\it Introduction} --		
%%%%%%%%%%%%%%%
Biological systems are often able to respond to small changes in the environment within a relatively short time. 
Conformational spread proposed for the lattice of E. coli chemoreceptors is a famous example, where the coupling strength between the activation states of nearest-neighbor receptors is close to a critical point such that small changes in the concentration of chemoattractant in the environment can be detected~\cite{ref:Duke_Bray_PNAS_1999}. 
In the cochlea, the amplification of weak sound is achieved by the response of a collection of motor proteins self-tuned near a Hopf bifurcation~\cite{ref:Duke_PNAS_2000}. 
It has also been suggested that the moving surface adhesive proteins in the gliding bacteria could be regulated close to a pitchfork bifurcation to enhance the response to small inhomogeneity of the environment~\cite{ref:Wada_Nakane_Chen_PRL_2013}. 
In these examples, energy is expended to keep the system close to a critical point, thereby enhancing its response to small variations in the signal (for E. coli chemotaxis) or to weak signals (for the cochlea and gliding bacteria) in a noisy environment. 

The activation of T cell receptors (TCRs) is another example with strong signal amplification: even a few agonist peptides can activate many T cell receptors in a very short time (typically less than 2 minutes)~\cite{ref:Eisen_Immunity_1996}.  
However, the mechanism of signal amplification for TCRs is not well understood.
It is known that 
TCRs form dynamic pre-clusters of size ranging from 30 to 300 nm, containing tens of TCRs before encountering agonist peptides~\cite{ref:Sammelson_FronCellDevBiol_2021}.
After forming stable bonds with agonist peptides, TCR-containing microclusters are crucial for T cell activation~\cite{ref:Saito_Nature_Immun_2005}.  
Furthermore, besides binding to agonist peptides, TCRs also bind weakly to self-peptides on the surface of antigen-presenting cells~\cite{ref:Davis_Sem_Immun_2007}.  
The enzyme-driven activation-deactivation steps serve as the kinetic proofreading processes to prevent TCR activation in the absence of agonist peptides~\cite{ref:McKeithan_PNAS_1995}.  
Nevertheless, experimental studies have also shown the importance of self-peptides in enhancing T cell responses to agonist peptides~\cite{ref:Germain_Immun_Rev_2003}.
  
    Besides being important in immunology, understanding the design principle for systems that can produce a strong response to the presence of only a few molecules could have practical applications.  
For this purpose, in this {\it letter}, we propose a minimal model that includes only the essential 
components and interactions of the TCR activation process, as mentioned previously.
In the case of T cells, an effective interaction arising from the elasticity of the cell membrane exists between two receptors, favoring them to be in the same free or bound state~\cite{ref:Goulian_Pincus_BPJ_1994, ref:Wu_Chen_PRE_2006, ref:Farago_SM_2015}.
In our model, this translates to an Ising-like effective binding-state coupling between nearby receptors. 
The system also contains many ``incorrect'' ligands that can form weak complexes with the receptors 
and none or one ``correct ligand'' that can form a stronger complex with a receptor. 
A simple kinetic proofreading scheme is incorporated to prevent receptor activation in the absence of the correct ligand.

Although this model system is far less sophisticated than T cell receptors in vivo, we find that even when a single correct ligand binds to a receptor, many receptors in the receptor cluster can be activated, 
thereby amplifying the response to a weak signal.  
Not only the range of parameter values over which amplification occurs,	but also the time it takes to achieve a strong response are comparable to those of in vivo studies of TCR activation~\cite{ref:Saito_book}.

\begin{figure}
\centering
\includegraphics[width=.9\linewidth]{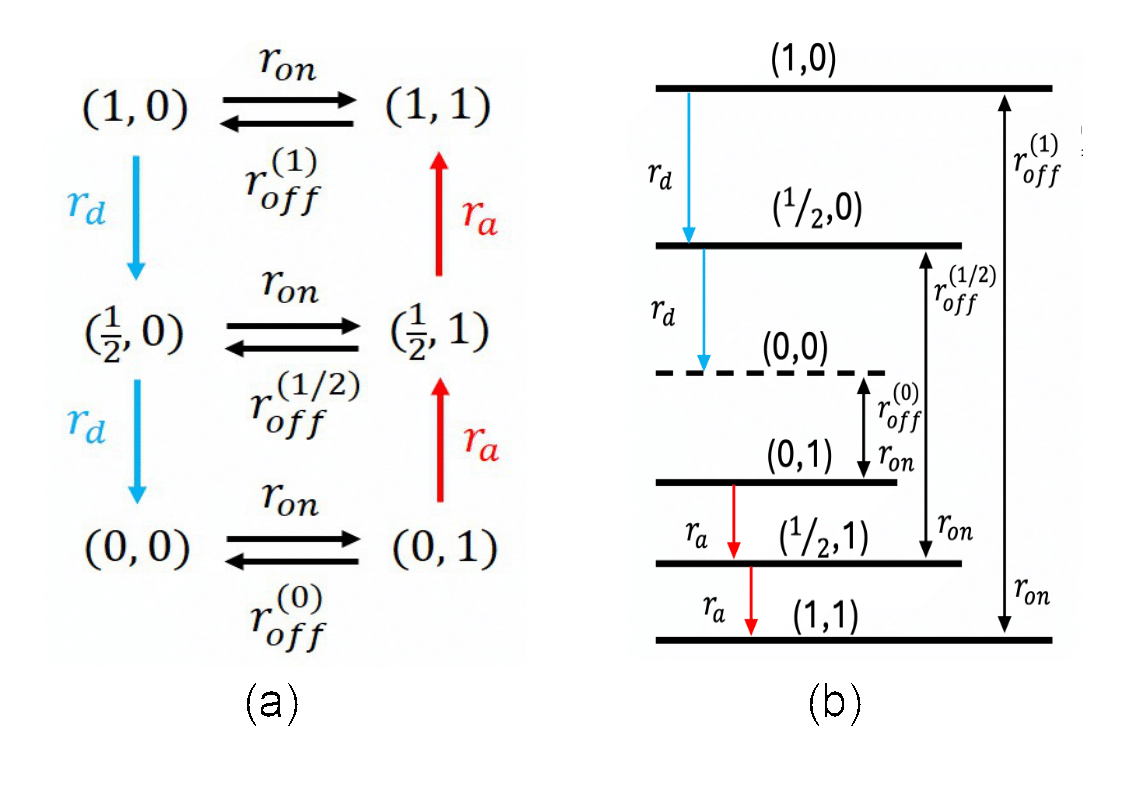}
\caption{
(a) The states $(a_i,b_i)$ of a receptor with $k=2$.  The bound states are shown on the right, the free states on the left. 
The binding rate is $r_{\rm on}$, the activation rate is $r_a$, and the deactivation rate is $r_d$. 
For simplicity, $r_{\rm on}$, $r_a$, and $r_d$ are constants.  
The unbinding rate $r^{(a)}_{\rm off}$, where $a = 0, 1/2, 1$, however, may depend on the activation level.
(b) The states of an isolated receptor arranged according to its energy.  
Activation of free receptors is forbidden; 
a bound-state receptor lowers its energy when it is activated.}
\label{fig:receptor_states}
\end{figure}

%%%%%%%%%%%%%%%%
{\it The model} --  
%%%%%%%%%%%%%%%%
     For simplicity, we define our model in two dimensions on a  $L \times L$ square lattice.  
Co-occupying the vertices are a square cluster of $N$ ($< L^2$) immobile receptors and two types of mobile ligands. 
A lattice site can be simultaneously occupied by multiple ligands, but a receptor can form a ligand-receptor complex with only one on-site ligand at a time. 
The ligands bind reversibly to the receptors, and the binding/unbinding kinetics obey detailed balance.   
An incorrect/correct ligand can form a weaker/stronger bond with a receptor as determined by different dissociation constants, 
otherwise there is no difference between these two types of ligands. 
Only ligands in the free state may diffuse freely with a diffusion constant $D$.
Let $b_i$ represent the binding state of the  $i$-th receptor ($i=1,2,..., N$).
$b_i=0$ when the $i$-th receptor is free, $b_i = + 1$ when it binds to an incorrect ligand, $-1$ to a correct one.  
In biology, a receptor often has a few activation levels, and the full activation of a receptor requires multiple activation steps~\cite{ref:Coombs_PLoSCompBiol_2009}.
In our model, a receptor can have its activation level $a_i = 0, 1/k, 2/k, \cdots, 1$.    
Only the receptors at activation level $1$ can transmit signals.    
The state of the $i$-th receptor is therefore characterized by $\left(a_i, b_i\right)$, as indicated in Fig.~\ref{fig:receptor_states}.  
In this {\it letter} we focus on the simple non-trivial case $k=2$. The generalization to other values of $k$ is straightforward. 

   The total energy of the system, including the binding-state coupling, is described by 
\begin{eqnarray}
    E &=& \sum_{i=1} ^N \left[(\epsilon_a - \Delta \epsilon \ {b_i}^2) \ {a_i} + \epsilon_b \ {b_i}^2 + \Delta {\epsilon_b} \ {b_i}\right] \nonumber  \\
 &&  - 4 J_b \sum_{\langle i,j \rangle} \left[\left({b_i}^2 - \frac{1}{2}\right)\left({b_j}^2 - \frac{1}{2}\right)\right],
\label{eq:energy}
\end{eqnarray}
where $\epsilon_a - \Delta \epsilon \, b_i^2$ is the activation energy for a receptor in binding state $b_i$, 
 $\exp{\left(\epsilon_b + \Delta {\epsilon_b} \right)}$ and $\exp{\left(\epsilon_b - \Delta {\epsilon_b} \right)}$ give the dissociation constant
of the incorrect and correct ligand, respectively,  
 $\sum_{\langle i,j \rangle}$ sums over all nearest-neighbor receptor pairs, 
 and $J_b$ is the strength of the coupling between the binding states of the nearest-neighbor receptors.
 
 In our model, the activation energy depends on the binding state of the receptor through $\Delta \epsilon$, it makes an active bound-state receptor less likely to unbind.  This choice mimics the 
 experimental finding that the lifetime of activated TCRs increases due to endogenous forces from the T cell~\cite{ref:Zhu_NatureComm_2023}, even though the catch bond property of the TCRs is not explicitly modeled.

 When a free receptor and a ligand occupy the same lattice site, a ligand-receptor complex can be formed with a rate $r_{\rm on}$. 
The unbinding rate $r_{\rm off}^{(a_i)}$ of receptor $i$ is related to $r_{\rm on}$ through the condition of detailed balance.
We further assume that the energy barrier between different activation states of a receptor is sufficiently high, such that the activation levels of a receptor can only be changed with the help of two types of specific enzymes: one type of enzyme activates the bound-state receptors with a rate $r_a$, 
and another type of enzyme deactivates the free-state receptors with a rate $r_d$.  
For simplicity, $r_{\rm on}$, $r_a$, and $r_d$ in our model are constants.  
When the conditions $r_a/r_{\rm off}^{(a_i)} < \mathcal{O}(1)$ and $r_d/r_{\rm on} > \mathcal{O}(1)$ are both satisfied, the ratio between the chance of an incorrect ligand-receptor complex becoming fully activated and that of a correct ligand-receptor complex can approach the $(k+1)$-th power of the limit set by the condition of detailed balance.  This kinetic proofreading mechanism is ubiquitous in biological systems~\cite{ref:Hopfield_PNAS_1974}. 

The receptor cluster, the incorrect ligands, and the enzymes that facilitate receptor activation/deactivation can be regarded as the constituents of a detector.  The strength of the detector's output signal is proportional to the number of receptors in the fully activated state.  A high-quality detector should have no fully activated receptors when no correct ligand is present, but in the presence of even a few correct ligand molecules, it should quickly reach a strongly signaling state.  

{\it Choice of parameters}  --  When a T cell is in contact with an antigen-presenting cell (APC),  a pre-cluster constains $\lesssim 30$ TCRs~\cite{ref:Sammelson_FronCellDevBiol_2021}, and a microcluster contains $30 - 300$ TCRs~\cite{ref_microclusters_number_first, ref_microclusters_number_second}. 
In the numerical simulations, our model detector is a $5 \times 5$ square cluster of receptors, comparable to a large pre-cluster or a small microcluster in a T cell.

The typical number of kinetic proofreading steps during the activation process of a TCR is estimated to be 5 to 10 steps~\cite{ref_kp_TCR}. We choose $k = 2$ for simplicity. 
Typical activation energy for a biological molecule is of the order of a few $k_{\rm B} T$,  
we choose $\epsilon_a = 3k_{\rm B} T$ and let the binding-induced decrease in activation energy be $\Delta \epsilon = 4 k_{\rm B} T$. 

The size of a T cell receptor is $\sim 10 \, \rm{nm}$~\cite{ref_receptor_size}. For convenience, we set $l_0 = 10 \, \rm{nm}$, the lattice constant and unit length in our model.
The unit energy is $k_{\rm B} T$.  

The dissociation constant $k_d^s$ of an incorrect ligand and $r_{\rm off}^{\left(0\right)}$ in the absence of the binding-state coupling are related to each other through the detailed balance condition
\begin{equation}
    k_d^s = \frac{r_{\rm off}^{\left(0\right)}}{r_{\rm on}} = \exp \left(\epsilon_b + \Delta \epsilon_b\right),
\end{equation}
  The unbinding rate of the self-pMHC is $\sim 3 \, \rm s^{-1}$, and the TCR-pMHC binding rate is $\sim 0.005 \, \rm \mu m ^2 \rm s^{-1}$~\cite{ref_lifetime_of_self_and_agonist}, this leads to 
  $k_d^s \sim 600  \mu {\rm m} ^{-2} \sim {\mathcal O}(0.01-0.1) l_0^{-2}$, therefore  
we choose $\epsilon_b + \Delta \epsilon_b = -1$. 
  
The unit time $t_0$ is chosen to make the inverse binding rate unity, {\it i.e.},  $r_{\rm on} = l_0^2/t_0$ and $r_{\rm off}^{(0)} = r_{\rm on} \exp(\epsilon_b + \Delta \epsilon_b) = e^{-1} t_0^{-1}$.
From the unbinding rate of the self-pMHC/TCR complex, the unit time $t_0$ in our model is $\sim 0.12 \rm \, s$. 

The difference of the binding energy between the correct and incorrect ligand, $\Delta \epsilon_b = 2.5 k_{\rm B}T$, is chosen by noting that the typical unbinding rate of the correct ligand is $\sim 0.02 \, \rm s^{-1}$~\cite{ref_Grove_T_cell_APC, ref_unit_time}.

For kinetic proofreading to work properly, the lifetime of a ligand-receptor complex should not be longer than that of the activation time; therefore, the activation rate in our numerical simulations is chosen as $r_a = 0.04 \ t_0^{_-1}$. 
Similarly, the deactivation rate is chosen to be $r_d = 0.4 \ t_0^{-1}$ so that the chance of rebinding before deactivation is small for the average ligand concentration used in our simulations. 
Converting to the SI unit gives $r_a \sim 0.4 \, s^{-1}$, and $r_d \sim 3.27 \, s^{-1}$, both are comparable to those of T cell receptors~\cite{ref_lifetime_of_self_and_agonist}.

The diffusion constant of a ligand is chosen as $D = 0.1 \rm \mu m^2 \rm s^{-1}$~\cite{ref_dissociation_constant}. The magnitudes of the parameters are listed in Table~\ref{table_parameters}.

\begin{table}
\caption{Definitions and parameters in our model.}
\vskip 0.25 cm
\begin{tabular}{l|l|l}
\hline
Physical meaning & Symbol & Value \\
\hline
\hline
Unit energy & $k_{\rm B}T$ & \\
Unit length & $l_0$ & 10 nm \\
Unit time &  $t_0$ & $0.12 $ s \\
Activation energy (free state)& $\epsilon_a$ & $3 \, k_{\rm B} T$\\
Binding-induced activation energy reduction %Activation energy reduction due to binding
                                                                                        & $\Delta \epsilon$ & $4 \, k_{\rm B} T$\\
Binding energy & $\epsilon_b$ & $-3.5 \, k_{\rm B} T$\\
Binding energy difference for $b_i = \pm 1$& $\Delta \epsilon _b $ & 2.5 $k_{\rm B}T$\\
Dissociation constants  & & \\%of a 
(single free correct ligand)            & $e^{\epsilon_b - \Delta \epsilon_b }$ & $e^{-6}$\\
(single free incorrect ligand) & $ e^{\epsilon_b + \Delta \epsilon_b }$ & $e^{-1}$\\
Binding rate & ${\it r}_{\rm on}$ & $1.0 \ l_0^2/t_0$ \\
Activation rate (bound state) & $\it {r}_a$ & $0.042 \ t_0^{-1}$\\
Deactivation rate (free state) & $\it {r}_d$ & $0.4 \ t_0^{-1}$\\
Unbinding rates 
(incorrect ligand) 
                               & $r_{\rm off}^{\left(0\right)}$ & $0.37 \ t_0^{-1}$\\
                               & $r_{\rm off}^{\left(1/2\right)}$ & $0.049 \ t_0^{-1}$\\
                               & $r_{\rm off}^{\left(1\right)}$ & $0.006 \ t_0^{-1}$\\
Diffusion constant of a ligand & $D$ & 0.1 $\mu \rm{m}^2/s$ \\
\hline
\end{tabular}
\label{table_parameters}
\end{table}

In the simulations, the receptor cluster is located at the center of the $L\times L$ lattice, and periodic boundary conditions are applied to the lattice.  
For each $J_b$ and $ c$, the simulations were carried out for 100 runs, each consisting of    $9 \times 10^5$ iterations with a time step $\Delta t = 2.7 \times 10^{-3}$, which corresponds to $\sim 290$~s in real systems.  
The binding-state coupling term for the receptors on the boundary of the receptor cluster is chosen such that its neighbor outside the cluster (where there is actually no receptor) is set to be $b=0$.  
This unfavors the formation of ligand-receptor bonds at the cluster boundary, mimicking the relatively greater distance between the T cell membrane and that of the antigen-presenting cell away from a pre-cluster or microcluster of TCRs due  to the absence of any TCR-pMHC bonds.    

\begin{comment}
In the absence of the correct ligand, the number of fully activated receptors in the steady state, $N_A^{(0)}$, increases as the concentration of the incorrect ligand increases.  
This is shown in Fig.~\ref{fig:Na-incorrect}. Systems with $N_A^{(0)} \geq 2$ are considered to fail the specificity criterion. 
%, they occupy the gray region of Fig.~\ref{fig:phase-diagram}.
\end{comment}

\begin{figure}
\centering
\includegraphics[width=\linewidth]{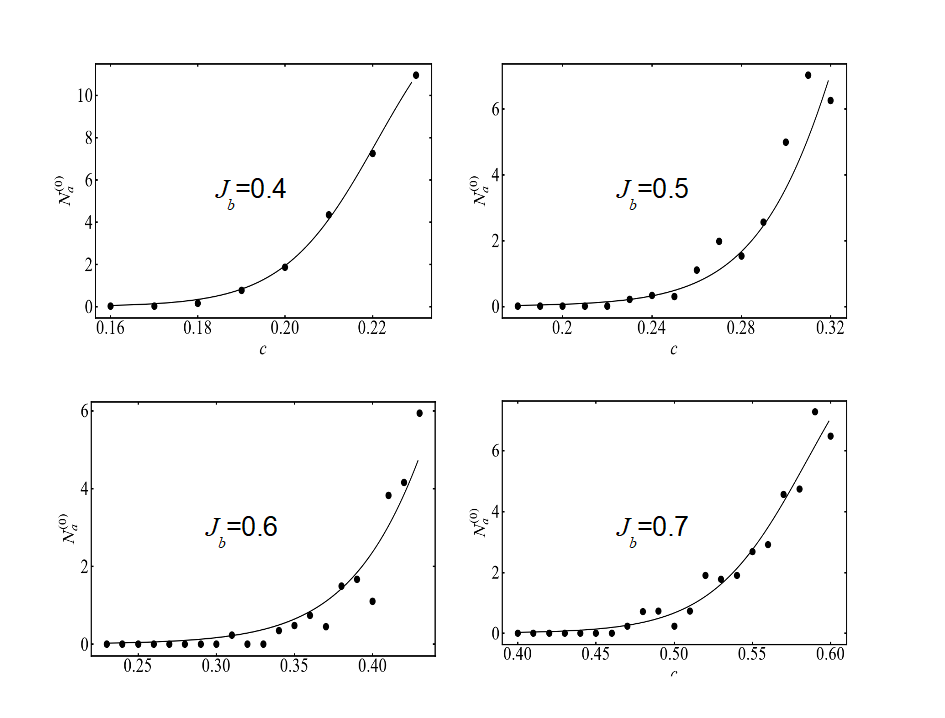}
\caption{$N_A^{(0)}$ at different $c$ for given $J_b$.  At sufficiently large $c$, $N_A^{(0)}$ becomes large, indicating failure of kinetic proofreading at large concentrations of incorrect ligands.
 }
\label{fig:Na-incorrect}
\end{figure}

%%%%%%%%%%%%%%%%%%%%%%%%%%%%%%%%%
{\it Signal amplification by a cluster of fixed receptors} --  
%%%%%%%%%%%%%%%%%%%%%%%%%%%%%%%%%
The performance of the detector in our model was studied using numerical simulations, with $L=21$, $ N=25$, and all other parameters summarized in Table~1. 

First, $c\times L^2$ incorrect ligands were randomly arranged in the lattice.  The system then evolved to a steady state $S_0$ with no correct ligand.  
Due to the kinetic proofreading mechanism, for a range of incorrect ligand concentrations $c$, there was no fully activated receptor in the system.  
As $c$ increases, however, some receptors can be activated 
(see Fig.~\ref{fig:Na-incorrect}). 
To avoid working at an unacceptably high concentration of incorrect ligands, we define a $J_b$-dependent threshold of incorrect-ligand concentration $c_{\rm max}(J_b)$: 
When $c > c_{\rm max}(J_b)$, the average number of active receptors at $t \approx 220 \ {\rm s}$, which we call $N_A^{(0)}$, is greater than $2$.

After the system reached the steady state $S_0$, a correct ligand was introduced to bind to the receptor at the center of the receptor cluster, and the system evolved toward a new steady state $S_1$. The evolution of the system was analyzed.     

\begin{figure}
\centering
\includegraphics[width=\linewidth]{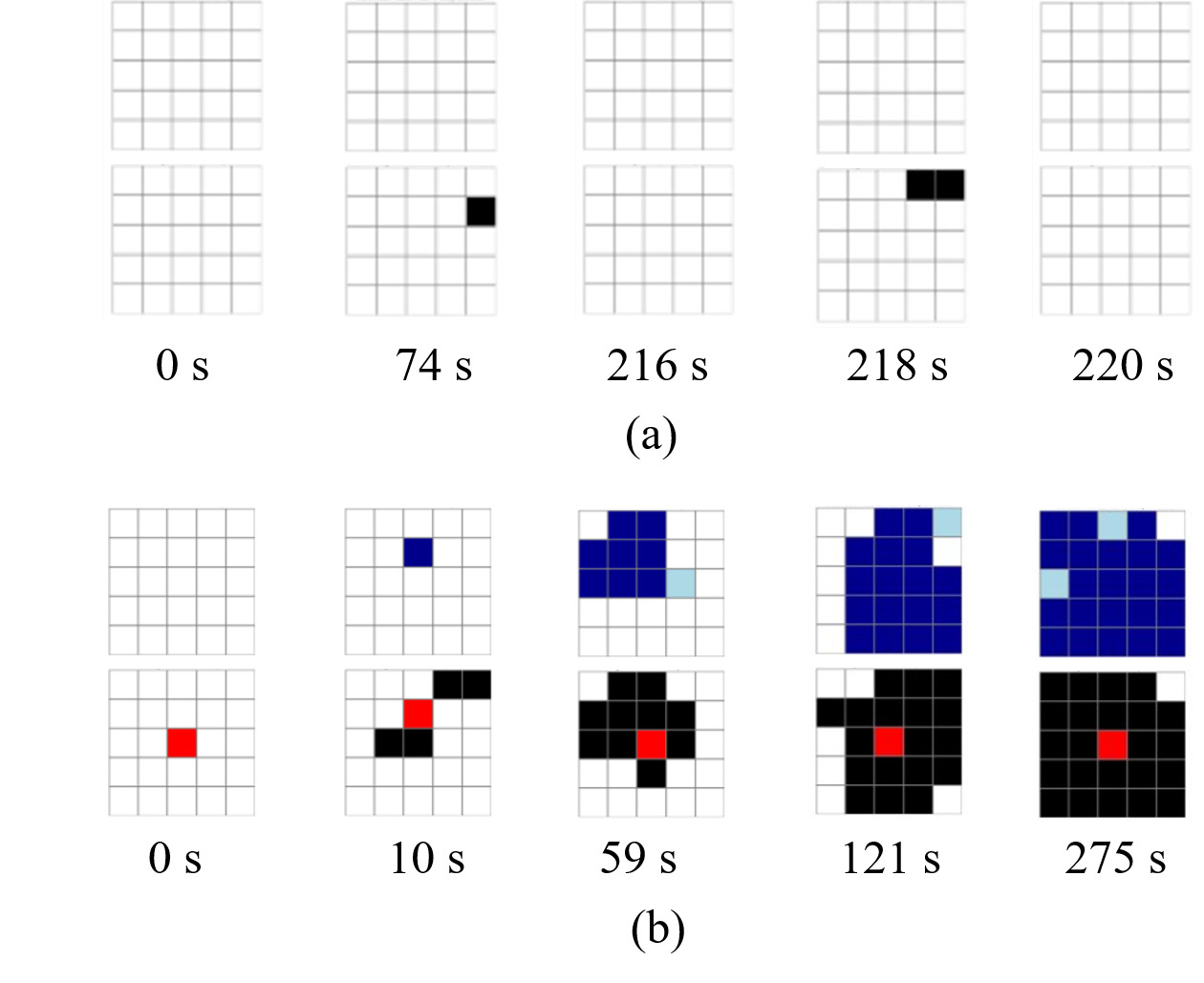}
\caption{Evolution of the activation state (white: $a=0$, light blue: $a=1/2$, dark blue: $a=1$) and binding state 
(white: $b=0$, black: $b=1$, red: $b=-1$) 
for a system without (a) and with (b) a correct ligand for concentration of incorrect ligand $c =0.32$ and binding-state coupling constant 
$J_b = 0.55$.  
Notice that during the simulation, the correct ligand binds to different receptors at different instances, indicating multiple unbind-diffusion-rebind events.  }
\label{fig:evolution}
\end{figure}

Figure~\ref{fig:evolution} shows the evolution of the activation state and binding state of the receptors in a typical simulation run without/with a correct ligand.  
In the absence of the correct ligand, although occasionally more than one receptor is in the bound state, no receptor becomes fully activated.  
On the other hand, a correct ligand acts as a nucleation seed for the incorrect ligands to form a stable bound state with the receptors due to the binding-state coupling.  
In this case, within a relatively short time interval, most of the receptors in the cluster are fully active because they form a stable bound state with incorrect ligands, even when the correct ligand unbinds and rebinds to different receptors a few times.

\begin{figure}
\centering
\includegraphics[width= 1.1 \linewidth]{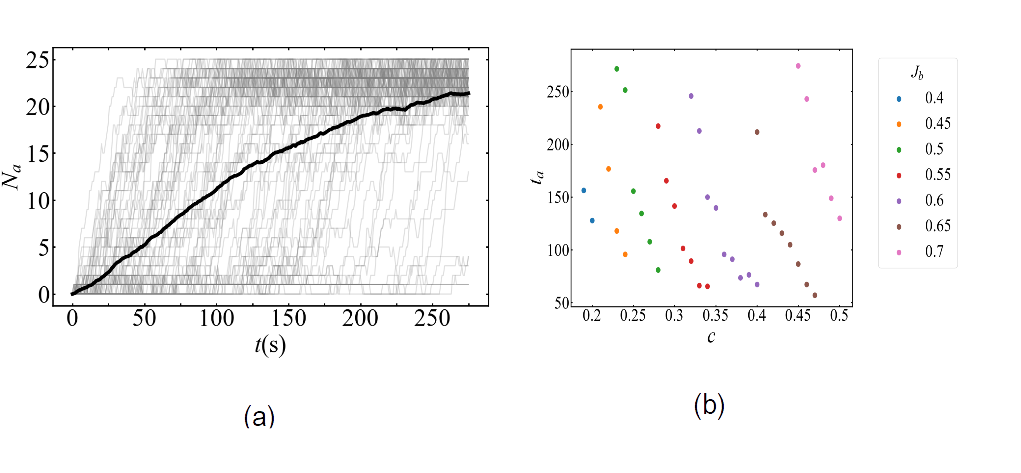}
\caption{(a) The number of fully activated receptors $N_a$ as a function of time $t$ in the presence of 1 correct ligand from 100 simulation runs for
$J_b = 0.55$, $c=0.32$.
The light gray lines are $N_a(t)$ from individual simulation runs; the black solid curve is the average over 100 runs.  
(b) $t_a$ versus $c$ at various $J_b$.  
 }
\label{fig:Na-vs-t}
\end{figure}

  The evolution of $N_a(t)$,  the number of fully activated receptors in the cluster at time $t$ after the correct ligand binds to the central receptor, 
shown in Fig.~\ref{fig:Na-vs-t}(a), exhibits strong sample-to-sample fluctuations due to the small $N$. 
To characterize the time the system needs to amplify the signal, we introduce $t_a$, the time at which the ensemble average of the number of fully active receptors reaches 10, i.e., when the signal is amplified 10-fold;   
we call $t_a$ the activation time.  
 As Fig.~\ref{fig:Na-vs-t}(b) 
 shows, at a given $J_b$, $t_a$ decreases as $c$ increases, and $t_a$ can be comparable to the activation time of the T cell receptors in a microcluster
~\cite{ref:Saito_Nature_Immun_2005}.  
Furthermore, when $J_b$ is small, because of weak receptor binding-state coupling, the smallest $t_a$ that a system can reach is large; when $J_b$ is large, the energy barrier between the free and bound states is high and $t_a$ is again large.  
These results suggest that the fast signal amplification requires a large $c$ within the range $c < c_{\rm max}(J_b)$; at the same time, $J_b$ cannot be too small or too large.

      To understand how $c$ and $J_b$ affect the signal amplification and activation time, it is instructive to note that by letting $s_i = 2 b_i^2 -1$ (i.e., a free receptor corresponds to $s_i = -1$, and a ligand-receptor complex corresponds to $s_i = +1$)
and $h = (\ln c -\epsilon _b - \Delta \epsilon _b)/2$, and if $a_i = 0$ for all receptors, the energy of the system becomes  
$- J_b \sum _{\langle i, j \rangle} s_i s_j - \sum _i h s_i $
when the correct ligand is not included.  
That is, in this situation, the receptor cluster can be mapped to a two-dimensional cluster of Ising ferromagnetic spins.  
Increasing $c$ is analogous to turning up the effective magnetic field $h$ in an Ising spin cluster.  
This increases the probability of collective spin flipping from $-1$ to $+1$ within the simulation time due to the mutual coupling $J_b$ and the cluster's finite size.  
Since $J_b$ also creates an energy barrier between $\sum _i s_i <0$ and $\sum _i s_i> 0$ macrostates, the collective spin-flipping time increases with $J_b$. 
The presence of a correct ligand-receptor complex corresponds to a spin cluster with an extra magnetic field $\Delta \epsilon _b$ at the site that the correct ligand-receptor complex occupies, favoring the transition from the $\sum _i s_i <0$ to the $\sum _i s_i >0$ macrostate.  
Therefore, the presence of a correct ligand-receptor complex encourages the incorrect ligands to move to the bound state collectively.  
Strong amplification requires a large $c$ with $c <c_{\rm max}(J_b)$, and a sufficiently large $J_b$ so that introducing the correct ligand-receptor complex can send the system to a macrostate with large $\sum _i {b_i}^2$.  
However, when $J_b$ is too large, the response time becomes too long for the system to perform well.

\begin{figure}
\centering
\includegraphics[width=.9\linewidth]{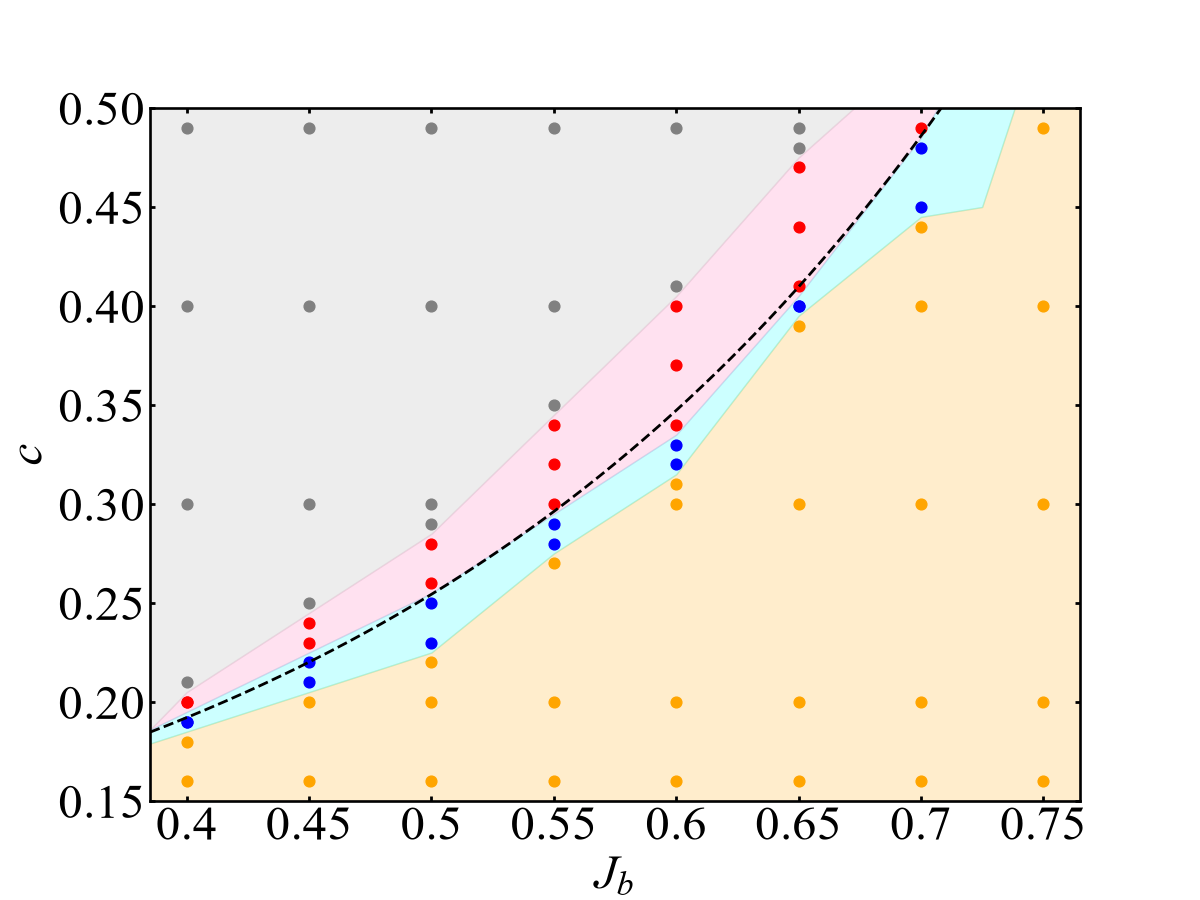}
\caption{ Phase diagram of our detector system.  The gray region is where kinetic proofreading fails.  In the light brown region, the correct ligand cannot reach $N_a > 10$ before $275$ s.  The light green region is where $275 \ {\rm s}>t_a > 150$ s, and the light red region is where $t_a < 150$ s. 
Data are collected at the locations marked by symbols. The dashed curve is the fitted function $c(J_b) = (81.9324 + 31.9463 \ e^{4J_b})t_0/t_a$ with $t_a = 150$~s. 
}
\label{fig:phase-diagram}
\end{figure}

   Overall, the detector's performance is determined by three factors.  
First, in the absence of any correct ligand, the receptors should not be activated; that is, $c < c_{\rm max}(J_b)$.  
Second, many receptors should be able to bind to the incorrect ligands after a correct ligand binds to a receptor in the cluster, i.e., $J_b$ and $c$ cannot be too small.  
Third, signal amplification should be achieved within a short time, i.e., $J_b$ cannot be too large.
   Fig.~\ref{fig:phase-diagram} shows that the 
light green and light red region where all these criteria are satisfied occupies a reasonably large range in the $c-J_b$ plane.   
The boundary between these two regions indicates where $t_a = 150$ s.  To reach  $N_a = 10$, a cluster of at least 10 bound receptors needs to form. 

In our model, activation lowers the energy of a receptor; it should proceed relatively fast compared to the formation of a cluster of bound-state receptors.   
Therefore, we postulate that the time to form a bound-state cluster of 10 receptors, $t_c$, largely accounts for the time 
for signal amplification to reach ten-fold. 
When all receptors are in the free state, growing a cluster of bound receptors requires forming an isolated bound receptor first, which has a high energy barrier.
After the correct ligand binds to a receptor, however, the barrier is lowered,
with the slowest process being ``flipping'' the binding state of an adjacent receptor which has 3 free and 1 bound nearest neighbors.  
This corresponds to an energy barrier $4 J_b$.  
(We neglect such cases where a bound-state receptor cluster emerges from merging two small clusters that grow out of two isolated bound receptors, as they
are rare and contribute very little to the weighted average $t_c$.)  
From this, one can easily guess that $t_c \approx t_0  (A_0 + A_4 e^{4J_b})/c$.  By setting $t_a = 150s$, one finds that $c(J_b) \approx (A_0 + A_4 e^{4J_b})t_0/t_a$, where $A_0$ and $A_4$ can be obtained by fitting to the midpoints between the red and blue points for a given $J_b$ in Fig.~\ref{fig:phase-diagram}. 
This simple estimate agrees very well with the simulations, indicating that the nucleation of a large bound-state receptor cluster is the limiting step of signal amplification. 

%%%%%%%%%%%%%%%%%%%%%%%%     
{\it Discussion and summary} --
%%%%%%%%%%%%%%%%%%%%%%%%
Our simulation results demonstrate that our simple model detector can achieve (i) high specificity, (ii) high sensitivity, and (iii) rapid response with kinetic proofreading and amplification of signal due to the prolonged lifetime of the incorrect ligand-receptor complexes in the presence of the correct ligand-receptor complex through the binding-state coupling between neighboring receptors.

Motivated by the phenomenology of TCR activation, our model detector shares several similarities with real biological systems.  
In particular, the binding-state coupling between receptors is motivated by the generic binding-state coupling observed in intercellular adhesion complexes. 
Naturally, as a minimal model, it is considerably simpler than a real system in several aspects: 
Although kinetic proofreading is an essential part of our detector, our model only considers receptors with three activation states, unlike receptors in immunological cells.  
Furthermore, our model detector is made of a cluster of immobile receptors on a two-dimensional lattice.  
Although TCRs form preclusters and microclusters in T cells, these clusters are dynamic, and TCRs in a precluster or a microcluster do not form a lattice~\cite{ref:Sammelson_FronCellDevBiol_2021,ref:Saito_Nature_Immun_2005}.

Despite the simplicity, our model captures the important role of incorrect ligands in signal amplification during TCR activation, a phenomenon not previously noted theoretically but exhibited in past experimental studies 
where the response of TCRs to incorrect ligands helps amplify the signal from TCRs~\cite{ref:Germain_Nature_2002}, 
although the detailed mechanism remains unclear~\cite{ref:Germain_Immun_Rev_2003}.  

In the literature, several immunological receptor-related models have been proposed to explain receptors' high sensitivity to only a few correct ligands.  The serial engagement model~\cite{ref:serial-engagement} requires a fine-tuning of the lifetime of a correct-ligand-receptor complex to achieve the activation of many receptors by only one ligand within a relatively short interval of time.  
The kinetic segregation model~\cite{ref:kinetic-segregation} assumes that T cell receptors in close contacts between a T cell and an antigen-presenting cell cannot be reached by large proteins that deactivate the receptors; therefore, bound receptors in these regions can be activated but free receptors cannot stay within before fully activated. 
More recently, it has also been proposed that a multivalent ligand, being able to bind to many receptors at the same time, can make the receptors sensitive to a very small number of ligands~\cite{ref:multivalent}. 
 Finally, when a receptor has many sites that can be activated after binding to a correct ligand, it is possible for this receptor to activate many downstream signaling molecules and let the immune cell produce a strong response~\cite{ref:Kueh_PNAS_2025}.  
These mechanisms rely on at least one of the following features: (1) a fine-tuned correct-ligand-receptor lifetime~\cite{ref:serial-engagement}, (2) a multivalent correct ligand~\cite{ref:multivalent}, (3) a receptor with many sites that can be activated~\cite{ref:Kueh_PNAS_2025}, and (4) size-segregation of bound ligand-receptor complexes and receptor deactivation enzymes~\cite{ref:kinetic-segregation}.    
While it is possible that in modern immune systems, one or more of those sophisticated mechanisms play a certain role, our model, on the contrary, relies only on simple features such as generic binding-state coupling and the presence of many incorrect ligands.  
Those features may be important when the immune cell response first appeared through biological evolution.

In summary, the good performance of our model receptor cluster across a realistic range of parameters sheds light on the basic mechanism underlying an important immunological response and also suggests a possible way to construct sensitive artificial detectors.  
The simplifications of our model compared to pre-clusters and microclusters of TCRs in vivo suggest room for improvements and future directions of studies about roles that the discovered mechanisms play in TCR activation.
        
{\it Acknowledgments} -- 
YRC and HYC thank the support of NSTC, Taiwan (Grant no.  NSTC 115-2112-M-008-005 -), and the National Center for  Theoretical Sciences (NCTS), Taiwan.
 
{\it Data availability} -- The data that support the findings of this study were generated by numerical simulations. The source code and parameters used to generate the simulations is publicly available~\cite{ref:data}.


\begin{thebibliography}{10}
\bibitem{ref:Duke_Bray_PNAS_1999}
T. Duke and D. Bray,
Heightened sensitivity of a lattice of membrane receptors.
Proc. Natl. Acad. Sci. USA, {\bf 96}, 10104 (1999).  

\bibitem{ref:Duke_PNAS_2000}
S. Camalet, T. Duke, F. J\"{u}licher, and J. Prost, 
Auditory sensitivity is provided by self-tuned critical oscillations of hair cells.
Proc. Natl. Acad. Sci. USA, {\bf 97}, 3183 (2002). 

\bibitem{ref:Wada_Nakane_Chen_PRL_2013}
H. Wada, D. Nakane, and Hsuan-Yi Chen, 
Bidirectional bacterial gliding motility is powered by the collective transport of cell surface proteins. 
Phys. Rev.. Lett., {\bf 111}, 248102 (2013). 

\bibitem{ref:Eisen_Immunity_1996}
Y. Sykulev, M. Joo, I. Vturina, T.J. Tsomides, H.N. Eisen, 
Evidence that a single peptide-MHC complex on a target cell can elicit a cytolytic T cell response. 
Immunity, {\bf 4}, 565 (1996). 

\bibitem{ref:Sammelson_FronCellDevBiol_2021}
L. Balagopalan, K. Raychaudhuri, and L.E. Samelson, 
Microclusters as T cell signaling hubs: structure, kinetics, and regulation, 
Front. Cell Dev. Biol., {\bf 8}, 608530 (2021). 

\bibitem{ref:Saito_Nature_Immun_2005}
T. Yokosuka, K. Sakata-Sogawa, W. Kobayashi, M. Hiroshima, A. Hashimoto-Tane, M. Tokunaga, M.L. Dustin, and T. Saito, 
Newly generated T cell receptor microclusters initiate and sustain T cell activation by the recruitment of Zap70 and SLP76.
Nature Immun., {\bf 6} 1253 (2005).  
T. Yokosuka, T. Saito, The immunological synapse, TCR microclusters, and T cell activation, in {\it Immunological synapse}, (Springer-Verlag, Berlin Heidelberg, 2010).

\bibitem{ref:Davis_Sem_Immun_2007}
M. Krogsgaard, J. Juang, and M.M. Davis, 
A role for ``self'' in T-cell activation.
Seminars in Immunology, {\bf 19}, 236 (2007). 

\bibitem{ref:McKeithan_PNAS_1995} 
T.W. McKeithan,
Kinetic proofreading in T-cell receptor signal transduction, 
Proc. Natl. Acad. Sci. USA, {\bf 92}, 5042 (1995). 

\bibitem{ref:Germain_Immun_Rev_2003}
I. \v{S}tefano\'{v}a, J.R. Dorfman, M.Tsukamoto, and R.N. Germain, 
On the role of self-recognition in T cell responses to foreign antigen, 
Immun. Rev. {\bf 191}, 97 (2003).

\bibitem{ref:Goulian_Pincus_BPJ_1994}
R. Bruinsma, M. Goulian, P. Pincus, Self-assembly of membrane junctions, Biophys. J, {\bf 67}, 746 (1994).

\bibitem{ref:Wu_Chen_PRE_2006}
JY Wu, HY Chen, Membrane-adhesion-induced phase separation of two species of junctions, Phys. Rev. E., {\bf 73}, 011914, (2006).

\bibitem{ref:Farago_SM_2015}
N. Dharan, O. Farago, Formation of semi-dilute adhesion domains driven by weak elasticity-mediated interactions, Soft Matter, {\bf 11}, 3780 (2015).

\bibitem{ref:Saito_book} 
T. Yokosuka and T. Saito, 
The Immunological Synapse, TCR Microclusters, and T Cell Activation, 
in T. Saito and F.D. Batista (eds.), Immunological Synapse, 
Springer-Verlag, Berlin Heidelberg, 2010. 


\bibitem{ref:Coombs_PLoSCompBiol_2009}
O. Dushek, R. Das, D. Coombs, 
A role for rebinding in rapid and reliable T cell responses to antigen, 
PLoS Comput. Biol., {\bf 5}, 1 (2009).

\bibitem{ref:Zhu_NatureComm_2023}
H-K. Choi, P. Cong, C. Ge, A. Natarajan, B. Liu, Y. Zhang, K. Li, M.N. Rushdi, W. Chen, J. Lou, M. Krogsgaard, and C. Zhu, 
Catch bond models may explain how force amplifies TCR signaling and antigen discrimination, 
Nature Commun., {\bf 14}, 2616 (2023).  

\bibitem{ref:Hopfield_PNAS_1974}
J. J. Hopfield, Kinetic proofreading: a new mechanism for reducing errors in biosynthetic processes requiring high specificity, Proc. Natl. Acad. Sci., {\bf 71}, 4135 (1974).
J. Ninio, Kinetic amplification of enzyme discrimination,
Biochimie 57, 587 (1975).

\bibitem{ref_microclusters_number_first}
T. Yokosuka, K. Sakata-Sogawa, W. Kobayashi, M. Hiroshima, A. Hashimoto-Tane, M. Tokunaga, ML Dustin, T. Saito, Newly generated T cell receptor microclusters initiate and sustain T cell activation by recruitment of Zap70 and SLP-76, Nat. Immunol, {\bf 6}, 1253 (2005).

\bibitem{ref_microclusters_number_second}
R. Varma, G. Campi, T. Yokosuka, T. Saito, ML Dustin, T cell receptor-proximal signals are sustained in peripheral microclusters and terminated in the central supramolecular activation cluster, Immunity, {\bf 25}, 117 (2006).

\bibitem{ref_kp_TCR}
T. W. McKeithan, Kinetic proofreading in T cell receptor signal transduction, Proc. Natl. Acad. Sci., {\bf 92}, 5042 (1995).

\bibitem{ref_receptor_size}
M. K. Wild, A. Cambiaggi, M. H. Brown, E. A. Davies, H. Ohno, T. Saito, P. Anton van der Merwe, Dependence of T Cell Antigen Recognition on the Dimensions of an Accessory Receptor–Ligand Complex, J Exp Med 190 (1) (1999).

\bibitem{ref_lifetime_of_self_and_agonist}
N. J. Burroughs, Z. Lazic, and A. van der Merwe, Ligand Detection and Discrimination by Spatial Relocalization: A Kinase-Phosphatase Segregation Model of TCR Activation, Biophys. J, {\bf 91}, 1619 (2006).

\bibitem{ref_exp_activation_time}
M. Fritzsche and K. Kruse, Mechanical force matters in early T cell activation, Proc. Natl. Acad. Sci., {\bf 121}, 37 (2024).

\bibitem{ref_Grove_T_cell_APC}
S. Y. Qi, J. T. Groves, A. K. Chakraborty, Synaptic pattern formation during cellular recognition, Proc. Natl. Acad. Sci., {\bf 98}, 12 (2001).

\bibitem{ref_unit_time}
M. N. Artyomov, M. Lis, S. Devadas, M. M. Davis, A. K. Chakraborty, CD4 and CD8 binding to MHC molecules primarily acts to enhance Lck delivery, Proc. Natl. Acad. Sci., {\bf 107}, 16916 (2010).

\bibitem{ref_dissociation_constant}
M. T. Figge and M. Meyer-Hermann, Modeling receptor-ligand binding kinetics in immunological synapse formation, Eur. Phys. J. D, {\bf 51}, 153 (2009).


\bibitem{ref:Germain_Nature_2002}
I. \v{S}tefano\'{v}a, J.R. Dorfman, and R.N. Germain, 
Self-recognition promotes the foreign antigen sensitivity of naive T lymphocytes, 
Nature, {\bf 420}, 429 (2002). 

\bibitem{ref:serial-engagement}
S. Valitutti, S. M\"{u}ller, M. Cella, E. Padovan, A. Lanzavecchia, 
Serial triggering of many T-cell receptors by a few peptide-MHC complexes, 
Nature, 375, 148 (1995). 

\bibitem{ref:kinetic-segregation}
P. A. Van Der Merwe and O. Dushek, Mechanisms for T cell
receptor triggering, Nat. Rev. Immunol. 11, 47 (2011).

\bibitem{ref:multivalent}
Z. Xie, S. Angioletti-Uberti, J. Dobnikar, D. Frenkel, and T. Curk, 
Receptor clustering tunes and sharpens the selectivity of multivalent binding, 
Proc. Natl. Acad. Sci., USA, {\bf 122}, e2417159122 (2025). 

\bibitem{ref:Kueh_PNAS_2025}
W.L. White, H.K. Yirdaw, A.J. Ben-Sasson, J.T. Groves, D. Baker, and H.Y. Kueh, 
Proofreading and single-molecule sensitivity in T cell receptor signaling by condensate nucleation.
Proc. Natl. Acad. Sci., USA, {\bf 122}, e2422787122 (2025). 

\bibitem{ref:data} https://github.com/YanRuChen-maker/Simulation-analysis

\end{thebibliography}
\end{document}